\documentclass[authoryear,preprint,review,12pt]{elsarticle}

\usepackage{amssymb}
\usepackage{amsmath, amssymb, geometry}
\usepackage{xcolor}
\usepackage[normalem]{ulem}
\usepackage{xurl}
\journal{Sustainable Cities and Society}
\usepackage{hyperref}
\hypersetup{
    colorlinks=true,
    citecolor=blue,
    linkcolor=blue,
    urlcolor=blue
}

\usepackage{setspace}
\usepackage{caption}
\begin{document}

\begin{frontmatter}

\title{Setting local road death targets through urban scaling}

\author[a,b]{Glaucia Guimarães Pereira}
\author[a]{Tálita Floriano dos Santos}
\author[b,c]{Diego Rybski}
\author[d]{Fabiano Lemes Ribeiro\corref{cor1}}
\cortext[cor1]{Corresponding author. E-mail address: fribeiro@ufla.br}

\affiliation[a]{organization={Sustainable Development Center, University of Brasília},
    addressline={Via L3 Norte, Campus Darcy Ribeiro},
    city={Brasília},
    postcode={70910-900},
    state={Distrito Federal},
    country={Brazil}}

\affiliation[b]{organization={Research Group Urban Complexity (UCP), Leibniz Institute of Ecological Urban and Regional Development (IOER)},
    addressline={Weberplatz 1},
    city={Dresden},
    postcode={01217},
    state={Saxony},
    country={Germany}}

\affiliation[c]{organization={Complexity Science Hub},
    addressline={Metternichgasse 8},
    city={Vienna},
    postcode={1030},
    country={Austria}}

\affiliation[d]{organization={Department of Physics, Federal University of Lavras},
    addressline={Trevo Rotatório Professor Edmir Sá Santos},
    city={Lavras},
    postcode={37200-900},
    state={Minas Gerais},
    country={Brazil}}

\begin{abstract}
\begin{singlespace}
Road injuries remain a leading cause of death worldwide. Sustainable Development Goal (SDG) target 3.6 calls for countries to halve road deaths by 2030, monitored through the rate of deaths per 100,000 inhabitants. Extending this rate-based metric to the local level implicitly assumes that road deaths vary proportionally with population, an assumption urban scaling theory suggests may not hold. We test this assumption and propose a model for setting local road death reduction targets that explicitly accounts for scaling effects. We analyze scaling patterns at the country level for United Nations member states in 2021, and at the local level for United States counties and Brazilian municipalities between 2010 and 2024. Road deaths scale linearly with population across countries, but sublinearly across local units in both Brazil and the United States, meaning that uniform rate-based targets are not appropriate, favoring larger units regardless of local policy performance. Using the Brazilian case, we estimated the model's policy parameter from the scaling exponent, which accounts for the scale advantage of larger cities and allocates the national reduction target across municipalities more fairly. These findings show that urban scaling, traditionally used to describe cities, can also be operationalized into policy design, translating national targets into locally fair commitments and offering a roadmap for global challenges that manifest locally.
\end{singlespace}
\end{abstract}

\end{frontmatter}

\section{Introduction}
\label{introduction}

Road injuries kill approximately 1.18 million people worldwide every year and are the leading cause of death among children and adults aged 5--29 \citep{who_global_2023, who_world_2025}. Road safety has been recognized as a public health problem since the early 20th century \citep{granados_increasing_2005}, and it is now included in the United Nations (UN) 2030 Agenda under Sustainable Development Goal (SDG) 3, Good Health and Well-Being.

SDG target 3.6 set the goal to ``by 2020, halve the number of global deaths and injuries from road traffic accidents'', monitored through indicator 3.6.1, the rate of deaths per inhabitant \citep{un_transforming_2015}. As road deaths declined by only 5.5\% from 2010 to 2021 \citep{who_world_2025}, the target was postponed to 2030 \citep{third_global_ministerial_conference_on_road_safety_stockholm_2020}, and further efforts are still needed worldwide to achieve it \citep{un_scope_2026, mohan_future_2021}. This slow progress suggests that the way road safety targets have been defined deserves closer scrutiny.

Although well-defined targets are a key factor in developing guidance and policies in health and epidemiology \citep{van_herten_targets_2000}, the UN proclaimed 2011--2020 the ``Decade of Action for Road Safety'' without one, simply inviting ``all Member States to set their own national road traffic casualty reduction targets to be achieved by the end of the Decade'' \citep{un_improving_2010}. Regional economic commissions in Africa and Asia-Pacific later proposed halving deaths \citep{un_improving_2012}, and the SDGs adopted it globally in 2015 \citep{un_transforming_2015}. 

SDGs are defined at the country level, but cities play a critical role in achieving many of their targets. Localization, the process of translating global policies and strategies to the local level \citep{valencia_adapting_2019, sever_sustainable_2025}, requires setting targets that are meaningful for cities themselves, not just scaled-down versions of national goals. This is not a trivial step. Comparing deaths per 100,000 inhabitants across countries, as indicator 3.6.1 does, is a convenient way to benchmark performance \citep{peden_world_2004}, but it assumes that the number of road deaths varies linearly with population. At the local level, this assumption may not hold, because cities of different sizes are not just scaled versions of one another \citep{bettencourt_growth_2007}.

Urban scaling theory \citep{rybski_urban_2025} formalizes this idea as an empirical regularity: many urban indicators do not grow linearly with population but follow power-law relationships instead \citep{bettencourt_origins_2013,ribeiro_mathematical_2023}. Under this framework, doubling a city's population does not simply double its outcomes. Due to increased interaction opportunities and network effects, socioeconomic indicators such as productivity and innovation tend to grow more than proportionally with city size. Conversely, undesirable outcomes such as crime and disease may also increase faster than population growth, meaning that larger cities can face disproportionately greater burdens \citep{bettencourt_growth_2007}. Infrastructural quantities, by contrast, such as road length and other physical networks, tend to grow more slowly than population, reflecting economies of scale in the provision of shared urban infrastructure \citep{bettencourt_origins_2013, kuhnert_scaling_2006}.

If road deaths follow a non-linear scaling pattern, then a linear reduction target, such as ``halving the number of deaths per inhabitant,'' may be too lenient for some cities and unrealistic for others. Therefore, setting local road safety targets without accounting for these urban scaling properties may lead to underperformance and misaligned policies.

Given the importance of well-defined targets for policy design, the cities' central role in achieving the SDGs, the need to localize targets, and the nonlinear nature of urban scaling, we propose a model for setting road death reduction targets that explicitly accounts for local scaling effects. To the best of our knowledge, no study has proposed using urban scaling for SDG localization. To address this gap, we proceed in three steps. First, we establish the worldwide scaling regime of road deaths at the country level. Second, we examine scaling regimes at the local level, using counties in the United States (USA) and municipalities in Brazil as sub-national units of analysis. Third, focusing on the Brazilian case, we propose a model for setting road death reduction targets that accounts for these scaling effects. 

\section{Materials and methods}

We developed a mathematical model grounded in urban scaling theory to set road death reduction targets across cities. In this section, we describe the concepts used regarding road deaths, local units, data, and urban scaling.

\subsection{Road death definition}

Road deaths are defined as ``fatalities caused by injuries sustained in traffic accidents involving one or more moving vehicles on public roadways'' \citep{who_definition_2025}. They may occur at the scene of the crash or later, as a consequence of the injuries sustained. Not all transportation deaths are road deaths, as deaths involving boats, trains, and aircraft are excluded.

Fatalities are recorded in death certificates under specific codes of the International Classification of Diseases (ICD). The SDG 3.6 metadata specifies a list of codes in the tenth revision (ICD-10), following World Health Organization (WHO) methods for country-level reporting \citep{who_who_2020} (see Section~\ref{sec:Data}).

\subsection{Local units}

The USA has an area of 9,158,022 km$^2$ and a population of 331 million \citep{us_census_bureau_census_2026}. Its 3,143 counties and county equivalents constitute the principal legal subdivisions of most of the 50 states, and also serve as the basic statistical units used in census data collection \citep{us_census_bureau_terms_2021, national_association_of_counties_what_2024}. Most counties function as governmental units, although their powers and responsibilities vary by state law. Road administration in the USA is distributed across multiple jurisdictional levels. Townships and municipalities are typically responsible for local streets and minor roads, counties often administer medium-scale road networks in unincorporated areas, and state governments retain authority over major highways and arterial roads \citep{levinson_hierarchy_2012}.

Brazil has an area of 8,509,361 km$^2$ and a population of 203 million. Its administrative structure includes 5,569 municipalities across 26 states,  plus the Federal District, which is counted as a municipality, totaling 5,570 municipal-level units \citep{ibge_cidades_2025}. Under the Brazilian Traffic Code, municipalities serve as the principal local traffic authorities, responsible for implementing, administering, and enforcing traffic policies at the local level \citep{brasil_codigo_1997}. Consequently, municipal governments formulate and implement many traffic management and urban mobility policies. As Brazil provides open-access data that follow the SDG 3.6 methodology (see Section~\ref{sec:Data}), it constitutes an ideal case for studying the urban scaling of road deaths.

Although they differ in formal administrative status, counties in the USA and municipalities in Brazil were treated here as functionally comparable local units, since both serve as the primary sub-national level at which population data are collected and, in Brazil, traffic policy is implemented. For simplicity, we refer to both as cities hereafter.

\subsection{Data}
\label{sec:Data}

All data are public and were processed in R, and the code is available upon request. Country-level data are from the Global Health Observatory \citep{who_global_2021} and geometries from the UN Geospatial Data service \citep{un_geospatial_united_2024}. For USA counties, road deaths come from the Fatality Analysis Reporting System \citep{nhtsa_fatality_2025} and were assigned to the county of occurrence, with population from the Census Bureau \citep{us_census_bureau_census_2026}.
For Brazilian municipalities, road deaths come from the Ministry of Health Mortality Information System, accessed through the \texttt{microdatasus} R package \citep{saldanha_microdatasus_2019}, and were assigned to the municipality of residence, with population from the \texttt{brpop} R package.  Road deaths were identified using the ICD-10 code ranges of the SDG 3.6.1 methodology \citep{who_who_2020}. Data were collected in August 2026.

\subsection{Urban scaling theory}

The current study is grounded in urban scaling theory, which posits that urban indicators $Y$ scale with the city population $N$ following a power law of the form $Y=Y_0N^\beta$ \citep{bettencourt_growth_2007,bettencourt_origins_2013,ribeiro_mathematical_2023}. Linearity occurs when $\beta \approx 1$, meaning that the urban indicator varies proportionally with population size. Otherwise, nonlinear regimes occur: when $\beta>1$, a superlinear regime occurs, indicating that the quantity increases more rapidly than population, which is the case for some socioeconomic urban metrics; when $\beta<1$, a sublinear regime occurs, meaning the quantity increases more slowly than population, which is the case for some infrastructure urban metrics.

Local-level scaling studies typically rely on administratively defined units, such as counties or municipalities, or on urban clusters delineated algorithmically from finer-grained statistical units through population density and commuting thresholds \citep{arcaute_constructing_2015}. Comparative analyses across multiple types of spatial delineations, ranging from administrative units to functional urban and metropolitan area definitions, have shown that scaling exponents can vary considerably depending on the choice of spatial unit \citep{cottineau_defining_2019, arcaute_constructing_2015}. Nevertheless, local units remain policy-relevant, particularly in Brazil, since municipalities are the administrative level responsible for implementing local traffic policy, making them a meaningful unit for setting road death reduction targets in practice.

\section{Results}

We pursued our analysis in three steps. First, we found a linear regime between the number of road deaths and population size at the country level and a sublinear regime at the city level (Figure~\ref{fig:country_local_level}). Second, we proposed an urban policy model based on these scaling properties. Finally, we estimated this policy for the Brazilian case quantitatively (Figure~\ref{fig:model_brazil}).

\subsection{Scaling at country level versus local level}

\begin{figure*}[!htbp]
    \centering
    \includegraphics[width=0.85\textwidth]{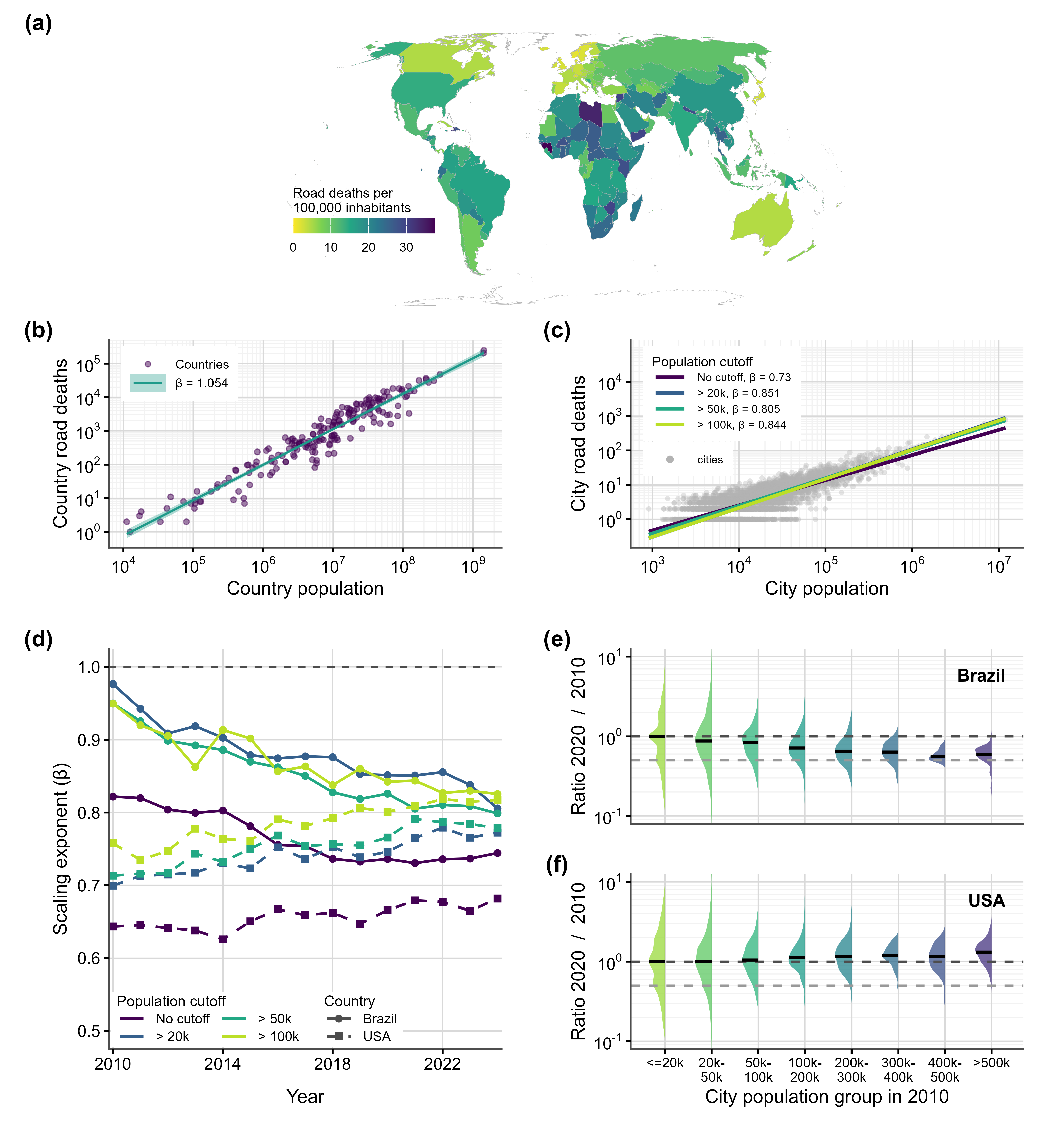}
\caption{Road death scaling regimes at the country and city levels.
(a) Road death rate per 100,000 inhabitants in 193 UN member states in 2021 (Equal Earth projection). Countries without data are shown in white.
(b) Road deaths versus population for the same UN countries (purple points), on log$_{10}$-log${10}$ axes. The green line is the ordinary least squares fit to the log-transformed data ($\beta = 1.054$, standard error = 0.022) and the shaded band is its 95\% confidence interval, indicating approximately linear scaling. (c) 
Power-law fits of the relationship between road deaths and population size for Brazilian cities in 2021, plotted on log${10}$-log${10}$ axes. Four population cutoffs are considered: all cities (no cutoff) and cities with more than 20,000, 50,000, or 100,000 inhabitants. Gray dots are individual cities.
(d) Scaling exponent $\beta$ by year from 2010 to 2024 for Brazilian (solid lines, circles) and USA (dashed lines, squares) cities, with the same cutoffs and colors as in (c). The horizontal dashed line marks linear scaling ($\beta = 1$). The exponent stayed below 1 (sublinear) for all years and all cutoffs in both countries. (e, f) Road death ratio (2020/2010) (log$_{10}$ axis truncated at 0.1 and 10) for (e) Brazilian and (f) USA cities, grouped by population size in 2010 (axes x). Distributions show kernel density estimates, and black lines show group medians. Gray dashed lines mark a ratio of 1 (no change) and 0.5 (SDG 3.6 target of halving road deaths). The median ratio falls with population in Brazil, but not in the USA, where distributions remain centered at or above 1.}
    \label{fig:country_local_level}
\end{figure*}

In 2021, 1,163,469 people died from road traffic injuries across 193 UN member states. Road traffic mortality rates per 100,000 inhabitants varied across countries (Figure~\ref{fig:country_local_level}a), from 1.3 in the Maldives and 1.5 in Norway to 34.0 in Libya and 37.4 in Guinea,  14.2 in the USA,  and 15.7 in Brazil. 

At country level, we performed a linear regression on a log–log scale, finding a power-law scaling pattern (Figure~\ref{fig:country_local_level}b). The estimated slope of 1.054 suggests that road deaths increase slightly more than proportionally with population size across countries. However, given the standard error of 0.022, we concluded that road deaths scale linearly with population under a power-law relationship. This means that comparing countries using rates per 100,000 inhabitants is an appropriate approach.

At the local level, we tested several population cutoffs to estimate $\beta$. Using Brazil in 2021 as an example (Figure~\ref{fig:country_local_level}c), city road deaths followed a power law with population. Using all cities gave the lowest exponent, whereas excluding cities with up to 20,000, 50,000, or 100,000 inhabitants gave fits that were close to each other. In all four cases, the exponent was below one. For the 100,000 cutoff, $\beta = 0.844$, meaning that a city ten times larger than another is expected to have approximately 30\% fewer road deaths per capita, since $10^{\beta-1} \approx 0.70$.

We then estimated the scaling exponent for each year from 2010 to 2024 (Figure~\ref{fig:country_local_level}d). In contrast to the country-level results, the exponents indicated a sublinear regime in both Brazil and the USA, showing that road deaths increase less than proportionally with city population. This sublinear pattern was robust to the choice of the cutoff used to exclude small cities. This means that comparing cities using rates per 100,000 inhabitants is not appropriate, because the number of deaths depends strongly on population size, with fewer deaths per 100,000 inhabitants in larger cities. Larger cities therefore have a scale advantage, which motivates a more intensive road death reduction policy for them, as proposed in the next section.

Between 2010 and 2020, during the first UN Decade of Action for Road Safety, road deaths in Brazil declined by 23.7\%, from 42,011 to 32,060, whereas in the USA they increased by 18.2\%, from 32,999 to 39,007. When disaggregating the ratio by population size in Brazil (Figure~\ref{fig:country_local_level}e), the median ratio decreased with population, indicating that the national decline was not uniform across cities, as larger cities achieved greater relative reductions in road deaths than smaller ones. In contrast, USA cities exhibited no substantial differences in the distributions of road death ratios across population groups, suggesting that the relative changes in road deaths were largely independent of population size. Moreover, the distributions remained centered around a ratio of 1 or larger, indicating little or no overall reduction in road deaths between 2010 and 2020. These findings suggest that road safety measures implemented in the USA during this period, if any, were insufficient to produce substantial reductions in road traffic fatalities, regardless of city population size.  (Figure~\ref{fig:country_local_level}f).

\subsection{The Model}

In this section, we propose a mathematical model for establishing road death reduction targets across cities, accounting for the nonlinear scaling regime. To begin, consider a country comprising $N_c$ cities, indexed by $i=1,\dots, N_c$. Let $P_i$ denote the population of the city $i$, and $Y_i$ its initial number of road deaths. The empirical scaling relationship between the number of road deaths, $Y_i$, and the population, $P_i$, can be expressed as

\begin{equation}
    Y_i = Y_0 P_i^{\beta} \, ,
    \label{eq:scaling}
\end{equation}
where $Y_0$ is the baseline coefficient and $\beta$ is the scaling exponent. For simplicity, we assume that the population $P_i$ remains constant throughout the implementation period of the reduction policy.

Our objective is to establish a target number of road deaths, denoted by $Y_i'$, for each city. To this end, we introduce a city-specific factor $\rho_i \in [0,1]$, which represents the city-specific effort and the fraction of the initial number of road deaths expected to remain after the implementation of the reduction policy. Accordingly, the target number of road deaths is given by
\begin{equation}
    Y_i' = \rho_i Y_i \, ,
    \label{eq:target_def}
\end{equation}
and therefore smaller values of $\rho_i$ correspond to more ambitious road death reduction targets.

To incorporate the scaling properties of road deaths into the policy framework, we assume that $\rho_i$ depends on population size according to a power-law relationship

\begin{equation}
    \rho_i = \rho_0 P_i^{- \lambda} \, ,
    \label{eq:rho_i}
\end{equation}
where $\rho_0$ is a normalization constant and $\lambda>0$ is a policy parameter that determines how the reduction effort is distributed among cities of different population sizes.
Substituting Equations~\eqref{eq:scaling} and~\eqref{eq:rho_i} into Equation~\eqref{eq:target_def}, we obtain
\begin{equation}
    Y_i' = \rho_0\,Y_0\, P_i^{\beta-\lambda} \, ,
    \label{eq:target_final}
\end{equation}
which establishes the scaling relationship between the target number of road deaths and population size.

We now impose the condition that the targets assigned to individual cities must collectively satisfy a predefined national road death reduction target. Let $Y^{tot}$ denote the total initial number of road deaths across all cities of the country, and $Y'^{tot}$ the corresponding total target number of deaths. We define $R\in[0,1]$ as the national reduction target, such that $R=0$ corresponds to no reduction and $R=1$ corresponds to the complete elimination of road deaths. Consequently, $R=0.5$ corresponds to a 50\% reduction in road traffic deaths, consistent with the target established under the SDG. The national target condition can therefore be expressed as
\begin{equation}
    Y'^{tot} =  \sum_{i=1}^{N_c} Y_i' =  \rho_0\,Y_0\, \sum_{i=1}^{N_c}  P_i^{\beta-\lambda}=(1-R)\,Y^{tot} \, .
    \label{eq:eq:sum_expanded}
\end{equation}

This condition allows us to find that $\rho_0 \propto (1-R)$, and consequently we obtain the population-dependent factor assigned to each city as

\begin{equation}
    \rho_i =  \left(\frac{Y^{tot}}{Y_0 \sum_{j=1}^{N_c} P_j^{\beta-\lambda}}\right)
    (1-R) P_i^{-\lambda} \, ,
    \label{eq:rho_i_solved}
\end{equation}
or simply

\begin{equation}
    \rho_i \propto (1-R) P_i^{-\lambda} \, .
    \label{eq:rho_i_solved_proportional}
\end{equation}

Thus, the proposed model establishes population-dependent road death reduction targets for individual cities while ensuring that the aggregate reduction satisfies the national target $R$.

Two reference values of $\lambda$ are particularly relevant. When $\lambda \approx 0$, all cities are assigned approximately the same factor $\rho_i \approx \rho_0$, regardless of population size. This corresponds to the conventional approach in which all cities are required to achieve the same proportional reduction in road deaths, without accounting for scaling effects.
In contrast, when $\lambda=\beta$, Equation~\eqref{eq:target_final} becomes independent of population size, implying that all cities are assigned the same absolute target number of road deaths.
Intermediate values of $\lambda$ between these two reference points represent different degrees of population-dependent adjustment in the allocation of reduction targets. The choice of $\lambda$ therefore reflects a policy decision regarding how reduction efforts should be distributed among cities of different sizes.

\subsection{Constraining the policy parameter}

Decision makers are free to choose $\lambda$ between 0 and $\beta$, but because well-defined targets are important \citep{van_herten_targets_2000}, we propose estimating the policy parameter $\lambda$ from urban scale theory. 

We used 2021 as the baseline year, in line with the SDG 3.6 reference, and Brazil to illustrate the exercise because road deaths declined in the country during the first Decade of Action. We chose a threshold of 100,000 inhabitants, yielding $\beta = 0.844$ (Figure~\ref{fig:country_local_level}c).

\begin{figure*}[!htbp]
    \centering
    \includegraphics[width=1\textwidth]{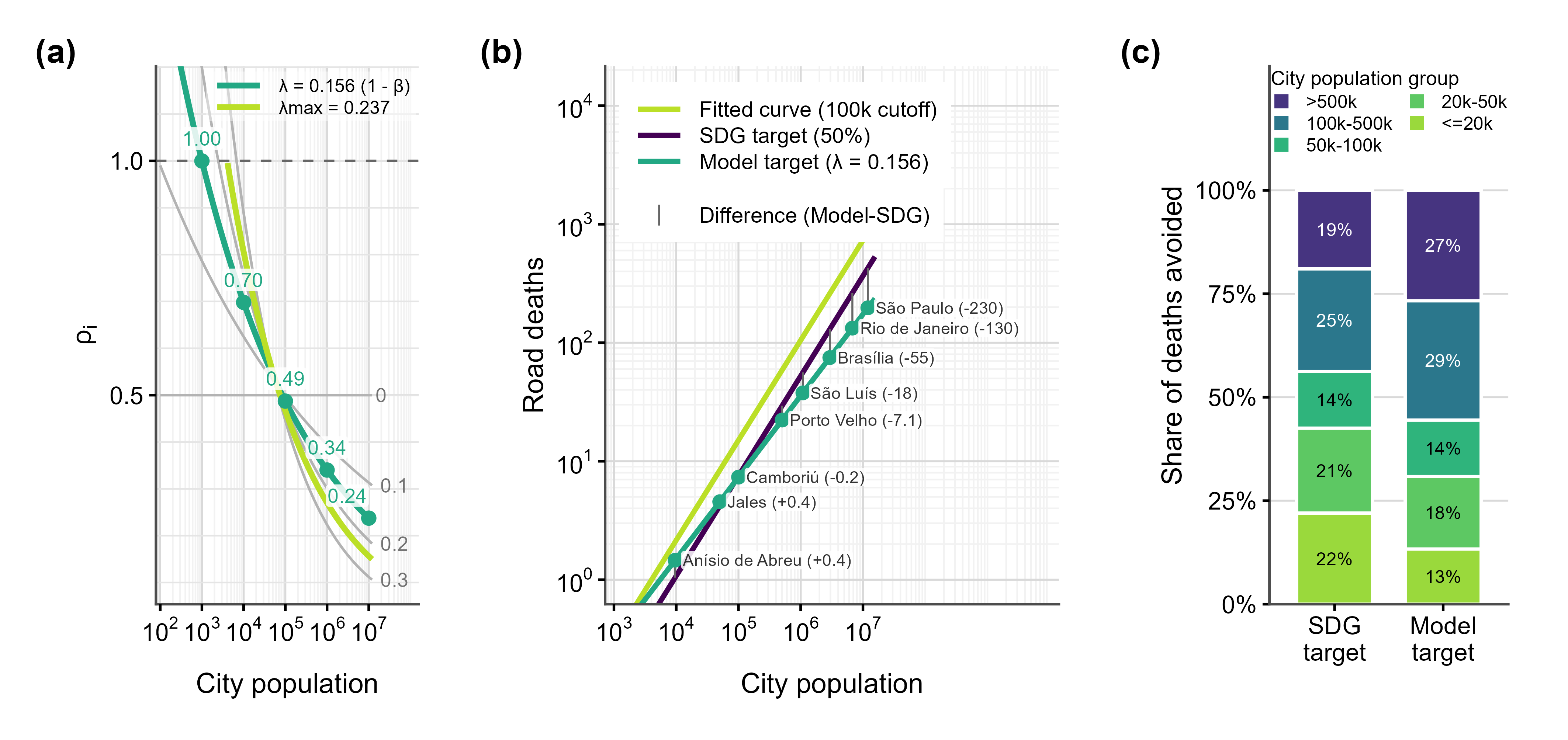}
    \caption{ Target-setting model applied to Brazilian cities in 2021.
    (a) Reduction factor $\rho_i$ as a function of city population (linear-log$_{10}$ scale), computed from $Y_0 = 9.07\times10^{-4}$ and $\beta = 0.844$. Gray curves show $\lambda \in \{0, 0.1, 0.2, 0.3\}$, the green curve shows $\lambda = 1 - \beta = 0.156$, and the yellow-green curve shows $\lambda = \lambda_{max} \approx 0.237$. The $\lambda_{max}$ curve is drawn only where $\rho_i \leq 1$, whereas the other curves are drawn over the whole range. Dots on the green curve mark its value at $10^3$ to $10^7$ inhabitants, wherever $\rho_i \leq 1$. The dashed horizontal line marks $\rho_i = 1$, above which a target would imply an increase in road deaths relative to the baseline.
    (b) Road deaths as a function of population (log$_{10}$ scale) for the  power law fitted with the 100,000 cutoff (yellow-green), the SDG 3.6 target ($R = 0.5$, purple) and the model-based target with $\lambda = 1 - \beta = 0.156$ (green). Eight Brazilian cities, selected to span different orders of magnitude, are placed on the model curve at their population. They are the three largest (São Paulo, Rio de Janeiro and Brasília), São Luís ($\sim$1 million inhabitants), and one city each of approximately 500, 100, 50 and 10 thousand inhabitants. Vertical gray segments show the gap between the two targets, and the labels give the difference in road deaths between the model target and the SDG target. The difference is negative for larger cities, which receive more demanding targets than under the SDG rule, and positive for smaller ones, which receive less demanding targets.  All values are read from the fitted curves and not from observed deaths.
    (c) Share of the national reduction in road deaths carried by each population band (up to 20,000, 20,000 to 50,000, 50,000 to 100,000, 100,000 to 500,000 and more than 500,000 inhabitants) under the SDG target and under the model target with $\lambda = 0.156$.    }
  \label{fig:model_brazil}
\end{figure*}

We first derived an upper limit for $\lambda$ that prevents any city from being assigned an increase in road deaths, narrowing the range of admissible values from $[0, \beta]$ to $[0, \lambda_{max}]$. As the national target is fixed, depending on the parameter $\lambda$ and population $P_i$, some cities’ targets $\rho_i$ can reach values greater than 1 (Figure~\ref{fig:model_brazil}a). The upper limit corresponds to setting $\rho_i = 1$ for the smallest city in the model. We defined this city as the one for which the fitted scaling law predicted a single death, which gave a population of $4{,}029$ inhabitants. Because the reduction factor cannot be inverted algebraically for $\lambda$, we found $\lambda_{max}$ numerically by root-finding. Applying the national reduction target $R = 0.5$ to the 2021 total $Y^{tot} = 33{,}471$, we obtained $\lambda_{max} \approx 0.237$, which is therefore an upper bound for the policy parameter.

To determine the policy parameter from scaling advantages in larger cities, we argue that small and large cities should face comparable levels of difficulty in achieving their road death reduction targets, while accounting for the structural advantages that larger cities already possess due to urban scaling. In particular, smaller cities may require less demanding reduction targets to compensate for their limited structural advantages, rather than necessarily receiving the same percentage reduction targets as larger cities.

The sublinear scaling relationship implies that road deaths per capita, namely $y_i \equiv Y_i/P_i$, decrease systematically with population size; that is, $y_i = Y_0 P_i^{\beta-1}$. Consequently, larger cities exhibit a scaling advantage in terms of road mortality. 

The question we address here is how this scaling advantage can be incorporated into the definition of city reduction targets. We therefore propose an alternative criterion based directly on the magnitude of the scaling advantage associated with urban scaling. To this purpose, consider two cities with populations $P_1$ and $P_2$, where $P_2 > P_1$. From the scaling relationship, the ratio between their expected road deaths per capita is
\begin{equation}\label{eq_y2_y1}
    \frac{y(P_2)}{y(P_1)}
    =
    \left(\frac{P_2}{P_1}\right)^{\beta-1}\, .
\end{equation}
Since $\beta < 1$, this ratio is smaller than one, reflecting the scaling advantage of the larger city.

On the other hand, the ratio between the reduction factors assigned to these cities is (from Eq.~\ref{eq:rho_i}) 
\begin{equation}\label{eq_rho2_rho1}
    \frac{\rho(P_2)}{\rho(P_1)}
    =
    \left(\frac{P_2}{P_1}\right)^{-\lambda} \, .
\end{equation}

We propose that the relative magnitude of the reduction factors should reflect the relative magnitude of the scaling advantage. Specifically, we impose
\begin{equation}\label{eq_imposition}
    \frac{y(P_2)}{y(P_1)}
    =
    \frac{\rho(P_2)}{\rho(P_1)}.
\end{equation}

Substituting the corresponding scaling relationships of the Eqs. \ref{eq_y2_y1} and~\ref{eq_rho2_rho1} into \ref{eq_imposition}  leads to
\begin{equation}\label{Eq_lambda}
\lambda = 1-\beta \, .
\end{equation}
For the Brazilian case, with $\beta = 0.844$, this criterion yielded $\lambda = 0.156$. This value was below the upper limit constraint of 0.237, indicating that the restriction was satisfied.

This choice  (Eq.~\ref{Eq_lambda}) establishes a direct relationship between the scaling exponent and the population dependence of city reduction targets. If road deaths scaled linearly with population ($\beta = 1$), there would be no scaling advantage associated with city size, and the model would assign the same percentage reduction target to all cities ($\lambda = 0$). Conversely, as the scaling relationship becomes increasingly sublinear, the value of $\lambda$ increases, assigning progressively more demanding percentage reduction targets to larger cities. This compensates for their scaling advantage by requiring stronger relative reductions. If every city achieves the reduction defined by $\lambda$, the scaling exponent becomes $\beta' = \beta - \lambda$ (Equation~\ref{eq:target_final} and Figure~\ref{fig:model_brazil}b).

As expected, the gap between the SDG 3.6 target and the model target in the largest cities showed that the proposed allocation asked considerably more of them (Figure~\ref{fig:model_brazil}b). For S\~ao Paulo, with approximately 12 million inhabitants, the model set a target 230 deaths lower than the SDG target. In contrast, for smaller cities such as An\'isio de Abreu (about 10 thousand inhabitants), the model target was higher than the SDG target, that is, it required a smaller reduction. As a consequence, cities with more than 500 thousand inhabitants accounted for 19\% of the national reduction under the SDG target and for 27\% under the model, whereas the smallest cities (up to 20 thousand inhabitants) went from 22\% to 13\% (Figure~\ref{fig:model_brazil}c).

\section{Discussion}
\label{discussion}

Our results show two distinct scaling regimes for road deaths, depending on the level of analysis. At the country level, road deaths scale approximately linearly with population, supporting the use of rates per 100,000 inhabitants for international comparisons, as currently adopted by SDG indicator 3.6.1. At the local level, however, both Brazilian and USA cities show a sublinear regime, meaning that road deaths increase less than proportionally with population size. In practice, this implies that road death rates tend to be lower in larger cities. So local-level comparisons are not appropriate, since they systematically favor larger cities regardless of actual policy performance.

This distinction is directly relevant for how road safety targets are currently set at the local level in Brazil. Several major Brazilian cities, including São Paulo, Rio de Janeiro, and Fortaleza \citep{camara_municipal_de_fortaleza_plano_2022}, have adopted road death reduction targets expressed as rates rather than absolute numbers. It is comparatively easier for large cities to reach the target independently of the effectiveness of local policy. Beyond this scaling inconsistency, rate-based targets are also difficult to reconcile with the Vision Zero paradigm, which treats every road death as unacceptable and rejects any target that implicitly tolerates a residual, non-zero level of mortality \citep{tingvall_vision_1999}. In this sense, a rate-based target can misguide policies away from this paradigm.

The common perception that large cities have more dangerous traffic leads to an expectation of superlinearity. Moreover, road deaths can be interpreted as forms of violence or crime \citep{bonilla-escobar_injuries_2014, shantajit_road_2018}, which usually scale superlinearly \citep{bettencourt_growth_2007}. Yet the results point in the opposite direction. Counterintuitively, proportionally fewer people died from road injuries in larger cities than in smaller ones. Thus, the mechanism that governs the scaling of road deaths differs from that governing other forms of urban violence.

One possible explanation is infrastructural. Road network extent has been shown to scale sublinearly with city population across different urban systems, with larger cities possessing proportionally less road infrastructure per capita than smaller ones \citep{louf_how_2014, arcaute_constructing_2015}. A smaller road network relative to population could, in principle, limit exposure in larger cities. However, the WHO lists the main risk factors identified as behavioral, including speeding, alcohol consumption, helmet use, seatbelt use, and distracted driving \citep{who_global_2023}. An infrastructural explanation is therefore, at best, a partial and indirect account of the observed pattern.

A more likely explanation lies in the composition of mobility patterns and speeds. Smaller cities tend to rely more heavily on motorcycles and less on public transportation, a combination associated with higher risk of severe or fatal injury. Larger cities, by contrast, tend to experience heavier congestion \citep{louf_how_2014, chang_is_2017}, which lowers average traffic speeds and can reduce fatalities even when the number of crashes remains high. Although posted speed limits, average travel speeds, and impact speeds are distinct quantities, they tend to be related \citep{elvik_updated_2019}. Lower speeds in larger cities, resulting from mobility patterns, can therefore help explain the sublinear relationship we found between road deaths and population size.

Scaling exponents are known to depend on how a city is defined, varying across administrative, functional, or algorithmically delineated units \citep{cottineau_defining_2019, arcaute_constructing_2015}. Here, this dependency is less of a concern, since the same units used to measure the scaling relationship also hold responsibility for local traffic policy, so the definition of a city and the definition of a policy-relevant unit coincide.
Another aspect that depends on definition is the geographic attribution of deaths. Because individuals may die in hospitals located in other cities, we assigned deaths to the deceased's city of residence \citep{ibge_indicadores_2024}. Using the place of death instead may lead to different scaling regimes \citep{rocha_non-linear_2015, sampaio_filho_evolving_2026}.

In this context, the combination of scaling effects, urban mobility aspects, and policy management suggests that larger cities have more room to reduce road deaths. The proposed model translates this into policy through a single parameter that shifts the reduction effort toward larger units while preserving the national target. Countries are responsible for putting this allocation into practice, since the target is national and its fair distribution across cities must be coordinated at that level. Setting $\lambda = 1 - \beta$, with $\beta$ estimated from the data, yielded a parsimonious allocation rule that depends on a single empirical parameter. Because $\beta$ is close to the value of about 0.85 reported for infrastructure-related quantities in urban scaling theory \citep{bettencourt_origins_2013, ribeiro_mathematical_2023}, the same rule could in principle be applied with $\lambda \approx 0.15$ as a reference value where local estimates are unavailable.

Some limitations should be noted. Cities are heterogeneous in territorial extension and mobility structure, and this heterogeneity is not explicitly modeled here. Our analysis focuses on the scaling relationship between population and road deaths and does not explicitly control for socioeconomic conditions, motorization levels, transportation mode shares, or the extent of road infrastructure. Despite these limitations, the consistency of the sublinear regime across two continental countries suggests that this pattern reflects a structural feature of road mortality at the local level, rather than an artifact specific to Brazil or the USA. Incorporating scaling effects into target-setting localization, as proposed in this study, offers a more coherent and equitable path toward reducing road deaths and advancing SDG 3.6 at the local level.

\section{Summary and conclusions}
\label{conclusion}

Road deaths remain a persistent burden for countries and cities. Despite recognition in the UN 2030 Agenda, more needs to be done. Because countries scale linearly while cities scale sublinearly, a pattern we confirmed across Brazilian municipalities and United States counties over more than a decade, larger cities have greater potential to reduce road deaths than smaller ones. We proposed a model that accounts for this pattern, using a single policy parameter to allocate the national target across cities more fairly.

Urban scaling theory has traditionally been used to describe urban patterns. In this study, we used it instead to improve a public policy, helping countries and cities reach their commitments.  This shows that urban scaling theory is useful beyond understanding cities, and can also be operationalized into policy design.

More broadly, the model may support the wider sustainability agenda, whose targets are typically set at the country level while many of the underlying problems arise at the city level. The same logic of translating a scaling pattern into policy targets could extend to other areas, providing a roadmap for countries to address global challenges that manifest locally.

\textbf{Declaration of generative AI and AI-assisted technologies in the manuscript preparation process.}

During the preparation of this work the authors used Claude AI to improve R codes. After using this tool/service, the authors reviewed and edited the content as needed and take full responsibility for the content of the published article.

\textbf{Acknowledgments}

G. G. Pereira thanks CAPES (grants number
88887.965945/2024-00 and 88881.218747/2025-01) for financial support, and the Leibniz Institute of Ecological Urban and Regional Development (IOER) in Dresden, Germany, for welcoming and hosting as a visiting PhD student.  F. L. Ribeiro thanks CNPq (grant numbers 403139/2021-0 and 424686/2021-0) and FAPEMIG (grant numbers APQ-00829-21 and APQ-06541-24) for financial support.

\textbf{Declaration of competing interests}

The authors declare that they have no known competing financial interests or personal relationships that could have appeared to influence the work reported in this paper.

\bibliographystyle{elsarticle-harv}

\begin{singlespace}
\renewcommand{\bibfont}{\footnotesize}
\setlength{\bibsep}{2pt}
\bibliography{ref_road_deaths}
\end{singlespace}

\end{document}